\documentclass{aa}

\usepackage[colorlinks=true,linkcolor=blue,citecolor=blue]{hyperref}
\usepackage{natbib}
\usepackage{siunitx}
\usepackage{xspace}
\usepackage{graphicx}
\usepackage{ulem}

\begin{document} 
  \title{Kinematic unrest of low mass galaxy groups}
  \author{G.~Gozaliasl \inst{1,2,3} \fnmsep \thanks{ghassem.gozaliasl@helsinki.fi}
     \and
           A.~Finoguenov \inst{2} 
           \and
           H.~G.~Khosroshahi \inst{4}
           \and
           C.~Laigle \inst{5}
           \and
           C.~C.~Kirkpatrick \inst{2} \and
           K.~Kiiveri \inst{2,3}
           \and
      J.~Devriendt \inst{6}
      \and
      Y.~Dubois \inst{5}
    \and
    J.~Ahoranta \inst{2}
    }
  \institute{Finnish centre for Astronomy with ESO (FINCA), Quantum, Vesilinnantie 5, University of Turku, FI-20014 Turku, Finland
         \and
         Department of Physics, University of Helsinki, P. O. Box 64, FI-00014 , Helsinki, Finland 
         \and
       Helsinki Institute of Physics, University of Helsinki, P.O. Box 64, FI-00014, Helsinki, Finland 
       \and
         School of Astronomy, Institute for Research in Fundamental Sciences (IPM), Tehran, Iran
         \and
          CNRS, UMR 7095 \& UPMC, Institut \v{d}Astrophysique de Paris, 98bis boulevard Arago, 75014 Paris, France
          \and
          Sub-department of Astrophysics, University of Oxford, Keble Road, Oxford OX1 3RH
     }
  \date{Received ; accepted }
\abstract
{In an effort to better understand the formation of galaxy groups, we examine the kinematics of a large sample of spectroscopically confirmed X-ray galaxy groups in the Cosmic Evolution Survey (COSMOS) with a high sampling of galaxy group members up to $z=1$. We compare our results with predictions from the cosmological hydrodynamical simulation of {\sc Horizon-AGN}. 
Using a phase-space analysis of dynamics of groups with halo masses of $M_{\mathrm{200c}}\sim 10^{12.6}-10^{14.50}M_\odot$, we show that the brightest group galaxies (BGG) in low mass galaxy groups ($M_{\mathrm{200c}}<2 \times 10^{13} M_\odot$) have larger proper motions relative to the group velocity dispersion than high mass groups. The dispersion in the ratio of the BGG proper velocity to the velocity dispersion of the group, $\sigma_{\mathrm{BGG}}/\sigma_{group}$, is on average $1.48 \pm 0.13$ for low mass groups and $1.01 \pm 0.09$ for high mass groups. A comparative analysis of the {\sc Horizon-AGN} simulation reveals a similar increase in the spread of peculiar velocities of BGGs with decreasing group mass, though consistency in the amplitude, shape, and mode of the BGG peculiar velocity distribution is only achieved for high mass groups.
The groups hosting a BGG with a large peculiar velocity are more likely to be offset from the $L_x-\sigma_{v}$ relation; this is probably because the peculiar motion of the BGG is influenced by the accretion of new members. }
\keywords{Galaxy groups--galaxies--Galaxy Cluster}
\maketitle

\section{Introduction} \label{sec:intro}

Galaxy groups represent a transitional environment between rich clusters and Milky Way-like halos. Understanding the dynamics of these structures is pivotal both for cosmology and galaxy evolution.
Unlike in galaxy clusters, scaling relations involving total gravitational mass, X-ray temperature, X-ray luminosity, group velocity dispersion, and other observable properties exhibit a large scatter \citep{khosroshahi07,mccarthy10,Wojtak2013}, which needs to be understood in order for galaxy groups to be considered as cosmological probes to the same level as clusters are. 

On the other hand, galaxy evolution depends on the assembly history of their host group. Indeed, the dynamical age of galaxy group halos has also been shown to be correlated with galaxy properties. For instance, \cite{Khosroshahi17} demonstrate that active galactic nuclei (AGN) radio flux, at a given stellar mass, is significantly lower for BGGs in dynamically relaxed groups compared to the brightest group galaxies (BGGs) of the same mass in dynamically evolving groups. This suggests that the AGN activity of BGGs, as probed by the radio emission, depends not only on the host mass but also on the dynamical state of the group (e.g. the degree of virialisation of the halo and the presence or absence of a second bright galaxy, as quantified from the luminosity gap).

A phase-space analysis of the group members should help to trace the assembly history of the groups and clusters of galaxies back. 
Seen in phase-space (the line-of-sight velocity versus the distance from cluster centre), galaxies which were accreted at early epochs do indeed tend to occupy the central virialised region with a low spread of relative velocities, while infalling or recently accreted galaxies have a higher relative velocity spread and are usually spatially offset from the centre of the virialised region \citep{noble2016phase}. The analysis of numerical simulations by \cite{rhee2017phase} shows that simulated galaxies tend to follow a typical path in phase-space as they settle into the cluster potential, and different regions of phase-space can be linked with different times since the first infall onto the cluster. From this analysis, they demonstrate that the location of cluster galaxies is connected to the tidal mass loss, hence quantifying how much galaxy evolution is impacted by the cluster environment.
\begin{figure*}
    \centering
    \includegraphics[width=0.7\textwidth]{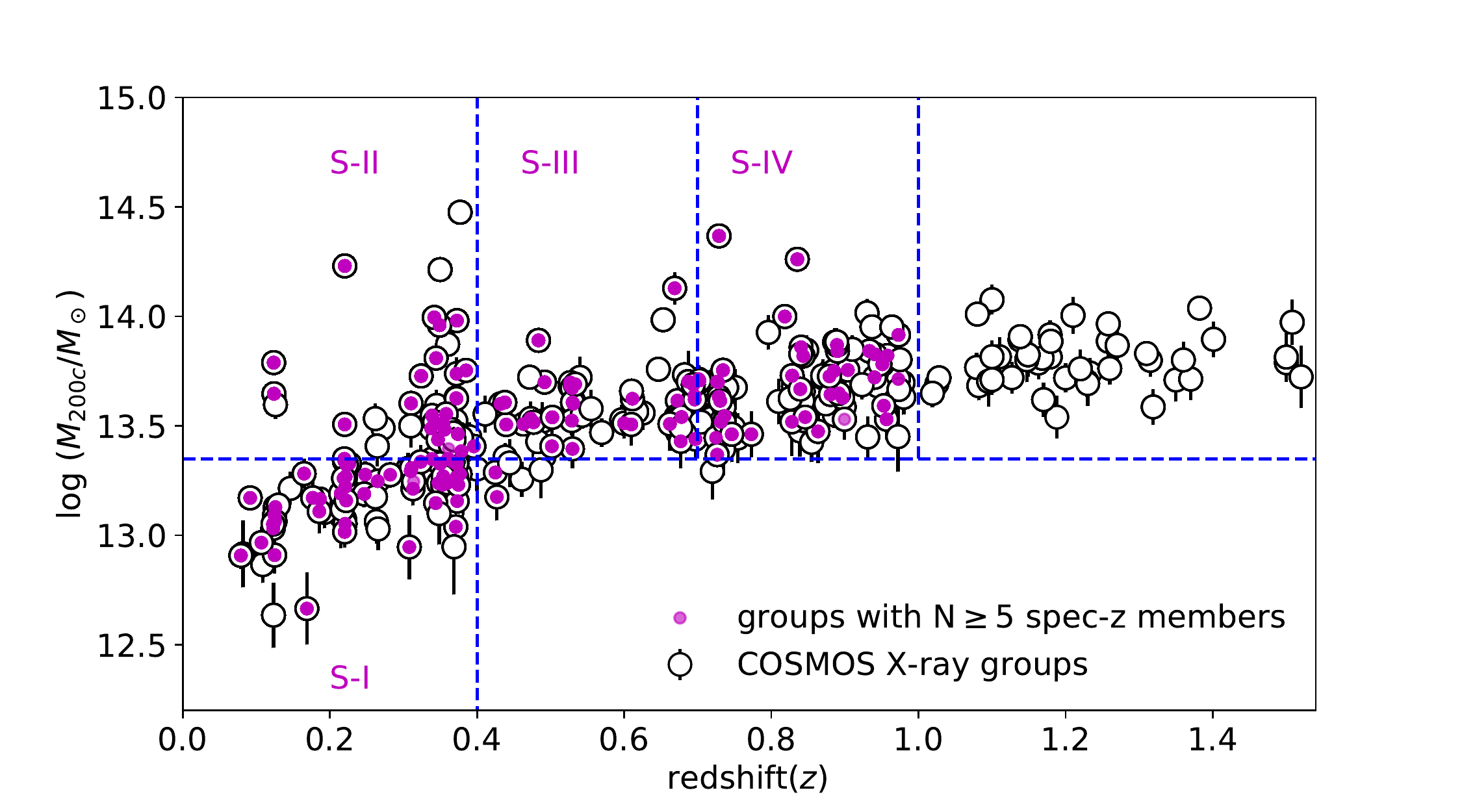}
    \caption{Distribution of the COSMOS X-ray groups in the halo mass versus redshift plane (open black circles). The filled magenta circles show the 95 groups with $N\geq5$ spectroscopic members which are used in this study. To inspect the redshift and halo mass evolution of the dynamics of BGGs and their satellites, we divide the hosting groups' sample (magenta circles) into four sub-samples, marked with four dashed blue boxes and labelled S-I, S-II, S-III, and S-IV.}
    \label{fig:m200_z}
\end{figure*}

In this work, we aim at applying similar techniques to a sample of X-ray selected galaxy groups in the Cosmic Evolution Survey (COSMOS) \citep{gozaliasl2019chandra} to shed light on their assembly history as a function of mass. COSMOS covers a two square-degree equatorial region of sky and was designed to probe the formation and evolution of galaxies as a function of the local galaxy environment and the cosmic time \citep{scoville2007cosmic}. The COSMOS field has been observed at all accessible wavelengths from the X-ray to the radio by several 
major space- and ground-based telescopes and offers a unique combination of deep (e.g. $ AB\sim 25-26 $ in the optical bands) multi-wavelength data. The COSMOS field has also been frequently targeted by several large spectroscopic programs, such as zCOSMOS, VIMOS Ultra Deep Survey (VUDS), FMOS-COSMOS, and Keck-DEIMOS \citep[see e.g.][]{lilly2007zcosmos,kartaltepe2010multiwavelength,comparat20150,hasinger2018}. We therefore rely on this wealth of spectroscopic follow up to confirm the membership of our group members and study the dynamics of the brightest group galaxies (BGGs) and their satellite galaxies since z=1.0 to the present day. 

This paper is organised as follows: section 2 describes the data, sample selection and the measurement of line-of-sight velocity and velocity dispersion. In section 3, we present the phase-space analysis and distribution of the relative peculiar velocity for satellites and BGGs, and the scaling relation between the X-ray luminosity and the observed velocity dispersion of groups (hereafter, $L_x-\sigma_{v, \mathrm{obs}}$).  We summarise the results together with our final remarks in section 4.
We assume a standard $\Lambda$CDM cosmology throughout the paper, with $H_{0}=70.4$ km $s^{-1} Mpc^{-1}$, $\Omega_\mathrm{M}=0.27,\;$ $\Omega_\Lambda=0.73.$
\section{Data and sample selection}

\subsection{ The COSMOS X-ray galaxy groups}

The COSMOS benefits from X-ray coverage by both the {\it Chandra X-ray Observatory} and {\it XMM-Newton}\footnote{For information on the COSMOS multi-wavelengths observations,  the list of broad-, intermediate- and narrow-band filters and filter transmissions, we refer readers to the COSMOS home web-page (\url{http://cosmos.astro.caltech.edu/}).}.
Several spectroscopic follow up campaigns have been carried out in the COSMOS field \citep[e.g.][]{lilly2007zcosmos,kartaltepe2010multiwavelength,comparat20150}. More recently, \cite{hasinger2018} presented a new catalogue of spectroscopic redshifts for 10,718 objects in COSMOS observed in the $550-980$ nm wavelength range using the Deep Imaging Multi-Object Spectrograph (DEIMOS) on the Keck II telescope.

The catalogue of X-ray galaxy groups used in this study has been presented in \cite{gozaliasl2019chandra}.
Once the redshift and group membership are estimated, a mass-dependent radial cut is chosen to sample analogous areas of each group. If the total mass of a group is known, the radial cut is determined using the following relation: 
\begin{equation}
M_{\Delta}=\dfrac{4\pi}{3} \Delta \times \rho_{crit}\times r_{\Delta}^3,
 \label{Eq:mh_r}
\end{equation}
where $\rho_{crit}$ is the critical density of the universe and $r_{\Delta}$ is the radius delimiting an interior density of  $\Delta$ times the critical density of the universe at the group redshift. In previous studies, $\Delta$ ranged usually between 180 and 500 times the mean or critical density in the Universe \citep{diaferio2001spatial,kravtsov2004dark}. In this study we assume  $\Delta=200$ and  apply $r_{200c}$ in our analysis.  

The halo mass of our groups are calculated from an empirical mass-luminosity
relation described in \cite{leauthaud2010weak} and applied to the COSMOS groups \citep[see also][]{connelly2012,kettula2015cfhtlens}: 
\begin{equation}
\begin{aligned}
\log_{10}(M_{200,c}) = p_0 - \log_{10}E(z) + \log_{10}(M_{0}) \\ 
+ p_1[\log_{10}(L_{x}/E(z)) - \log_{10}(L_{0})],
\end{aligned}
\label{Eq:m200}
\end{equation}
where $M_{200c}$ is the mass within $r_{200c}$, in units of ${\rm M}_{\odot}$.
 $p_0$ and $p_1$ 
are the fitting parameters, $
\log M_0$ and $ \log L_0$ are the
calibration parameters and $E(z)$ is the correction for the redshift evolution
of scaling relations. 
An extra error of $0.08$ dex  which corresponds to log-normal scatter in the $ L_x-M_{200c} $ relation is also included in our mass measurement as detailed in \cite{allevato2012occupation}. 
While we describe the sample using this convention, we also reexamine the relation of $L_x$ to the halo mass as traced by galaxy dynamics in \S\ref{SubSec:Lx-sigma}.

Fig.~\ref{fig:m200_z} shows the halo mass $\log_{10}(M_{\mathrm{200c}}/{\rm M}_{\odot}$) as a function of redshift for the entire sample of X-ray galaxy groups (open circles) in the COSMOS field. Halo masses ($M_{\mathrm{200c}}$)  range between $10^{12.5}$ to $10^{14.5}{\rm M}_{\odot}$ over  $0.07<z<1.53$. We highlight in magenta the 95 groups with $N\geq5$ spectroscopic members (excluding the BGGs). In order to study mass and redshift evolution, we define the following four sub-samples (labelled as S-I, S-II, S-III, and S-IV  in Fig.~\ref{fig:m200_z}):\\
$0.07<z<0.4$ \& $12.65<\log_{\mathrm{10}} (M_{\mathrm{200c}}/M_\odot)<13.30$ (S-I)\\
$0.07<z<0.4$ \& $13.30<\log_{\mathrm{10}}(M_{\mathrm{200c}}/M_\odot)<14.50$ (S-II)\\
$0.40<z<0.7$ \& $13.30<\log_{\mathrm{10}}(M_{\mathrm{200c}}/M_\odot)<14.50$ (S-III)\\
$0.70<z<1.0$ \& $13.30<\log_{\mathrm{10}}(M_{\mathrm{200c}}/M_\odot)<14.50$ (S-IV).

Throughout this paper, we use the position of the X-ray emission peak obtained by high spatial resolution Chandra imaging as a proxy for the group centre. This study also relies on our previous identification and selection of the BGGs \citep{gozaliasl2014gap,gozaliasl2019chandra}. In brief, the COSMOS2015 photometric redshift catalogue \citep{laigle2016cosmos2015} is used to rank galaxies as a function of mass, and in each group, a BGG is selected as the most massive galaxy. Groups for which a putative BGG does not have a spectroscopic redshift are not considered in this study.

In order to compute the observed velocity dispersion (hereafter, $\sigma_{v, \mathrm{obs}}$), we first select member galaxies for each cluster and group. Thanks to the wealth of COSMOS data, the redshift of each halo can be robustly estimated. The proper velocity of each galaxy within $r_{\mathrm{200c}}$ is first estimated from $v_{\rm prop} = c(z_g - z_h)/(1 + z_h)$ \citep{danese1980}, where $z_g$ and $z_h$ are the redshifts of the galaxy and its associated group halo, respectively. The velocity dispersion is then computed and galaxies deviating by more than 3-sigma are removed from the sample.  The groups are then visually inspected to remove additional outliers and substructure along the line of sight following the procedures described in \cite{clerc2016}.

We compute the mean redshift of the halo using the biweight average of the spectroscopic members \citep{beers90}, excluding the BGG.  The proper velocity, $v_{\rm prop}$, is recomputed for every galaxy using this redshift.  When a large number of members is available, the velocity dispersion, $\sigma_{\rm v, \mathrm{obs}}$, is calculated as the square root of the biweight variance of the member galaxies' proper velocity.  When the groups have less than 15 spectroscopic members, which occurs frequently in our sample, we use the gapper method (still excluding the BGG), known to give more robust results with a low number of members \citep{beers90}.

Following \cite{carlberg1997average}, the velocity dispersion of a group is estimated from the virial theorem (VT) as 
\begin{equation}
   \sigma_{v, \mathrm{VT}}=\dfrac{10\;r_{\mathrm{200c}}\times H(z)}{\sqrt{3}},
   \label{Eq.1}
\end{equation}
 where $H(z)$ is the Hubble constant at redshift $z$ and $r_{\mathrm{200c}}$ is the projected
and empirically determined radius of the group, the radius at which the mean interior
overdensity is 200 times the critical density.  In the simulation, we simply take the halo virial radius and convert it to $r_{200c}$  \citep[see e.g.][]{White2001}.

\subsection{The Horizon-AGN simulated light-cone}

In order to compare our observational results with theoretical predictions, we extracted a group catalogue from the hydrodynamical simulation light-cone of {\sc Horizon-AGN} \citep{dubois14}. 
The {\sc Horizon-AGN} is a cosmological hydrodynamical simulation (100 Mpc$/h$ a side) run with the adaptive mesh refinement (AMR) code {\sc RAMSES} \citep{teyssier02}, using a cosmology compatible with WMAP-7 \citep{Komatsu11}. The volume contains $1024^3$ dark matter (DM) particles (which corresponds to a DM mass resolution of $M_\mathrm{DM,res}=8\times 10^7 \, {\rm M}_\odot$). 
The  evolution of the gas is followed on the AMR grid down to a scale of 1~kpc, and includes gas heating by a uniform UV background \citep{Haardt1996} and cooling via H, He and metals \citep{sutherland&dopita93}. Star formation is modelled via a Schmidt law with a constant star formation efficiency per free-fall time of 2~percent \citep{kennicutt98}. Feedback from stellar winds and  supernovae (both type Ia and  II) is accounted for with mass, energy, and metal releases in the ambient inter-stellar medium. Feedback from black holes is  accounted for in either quasar or radio modes depending on the accretion rate. More details on the physics implemented in the simulation can be found in \cite{dubois14}. The simulation reproduces the overall evolution of galaxy populations throughout cosmic time \citep[see e.g.][]{dubois16,kaviraj2017horizon}.

The light-cone in the {\sc Horizon-AGN} box subtends 1 degree by 2.5 degrees out to redshift one. The evolution of the lightcone is sampled 22,000 times out to z=8.

The \textsc{ AdaptaHOP} halo finder \citep{aubert04} has been run on both the stellar and DM particle distributions in order to identify galaxies and halos respectively \citep[see][for more details]{laigle19}. For galaxies, local stellar particle density is determined from the 20 nearest neighbours, and structures are selected with a density threshold equal to 178 times the average matter density at that redshift. Only galaxies with more than 50 stellar particles (i.e., with $\log M_* > 10^{8}{\rm M}_{\odot}$) are kept in the catalogue. For halos the methodology is the same but with a density threshold of 80 times the average matter density. Halos with more than 100 DM particles are kept in the catalogue. As in \cite{darragh-ford19}, each galaxy is associated with its closest main halo. To match the observational definition, the BGG is identified as the most massive galaxy within the virial radius of the main halo.
Using Eq. \ref{Eq.1} introduced above, we obtain the velocity dispersion of the simulated groups using the virial mass of the hosting DM halo. As in observations, we refer to this estimation as $\sigma_{v,\mathrm{VT}}$. To match the observational limitation, the velocity dispersion of galaxy groups from the simulation is computed along one axis. We refer hereafter to the velocity dispersion of galaxies from the {\sc Horizon-AGN} simulation as $\sigma_{v,\mathrm{D1}}$. The choice of the axis for the projection does not impact our results.

\begin{figure}
    \centering
    \includegraphics[width=0.5\textwidth]{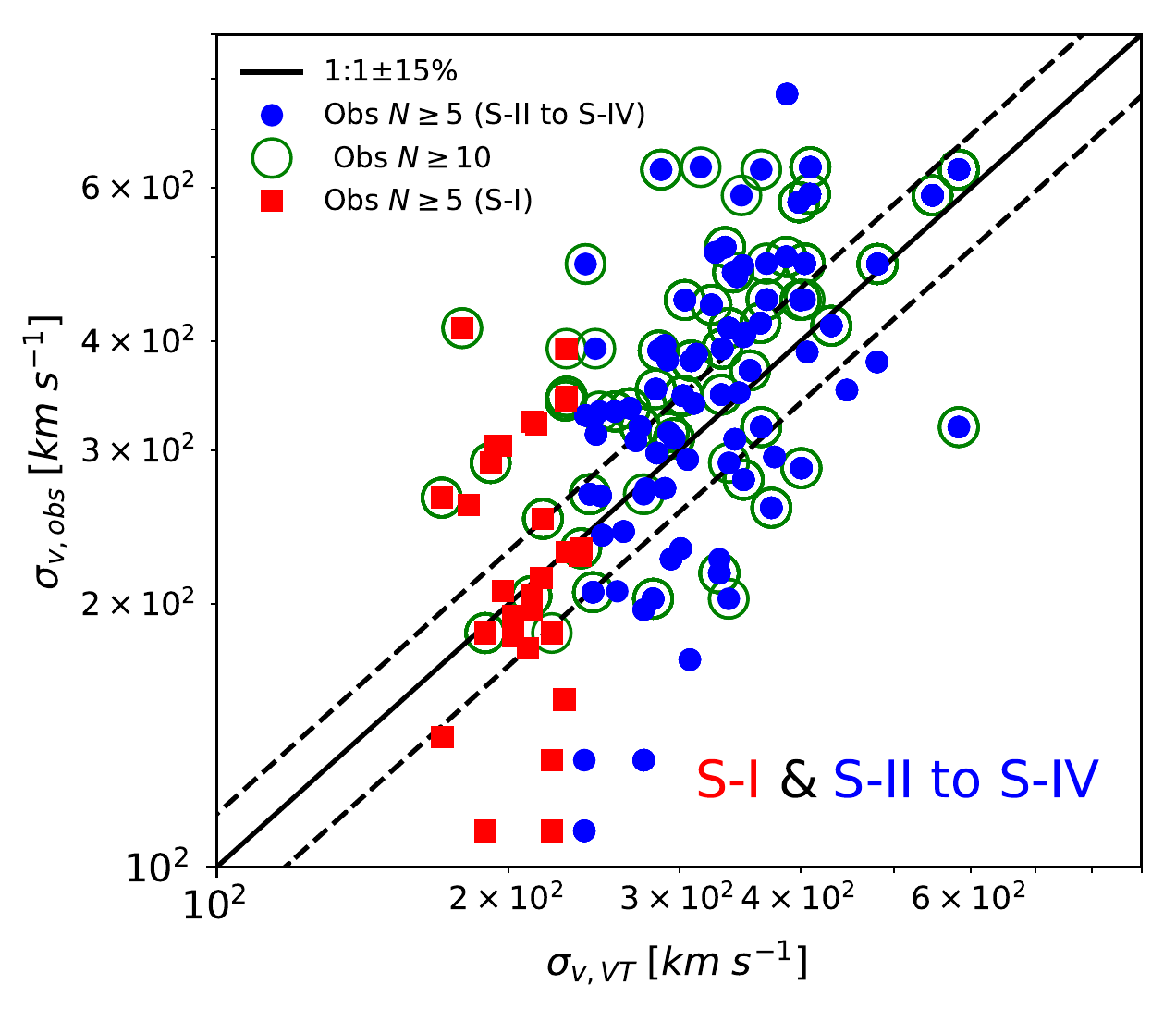}
        \includegraphics[width=0.5\textwidth]{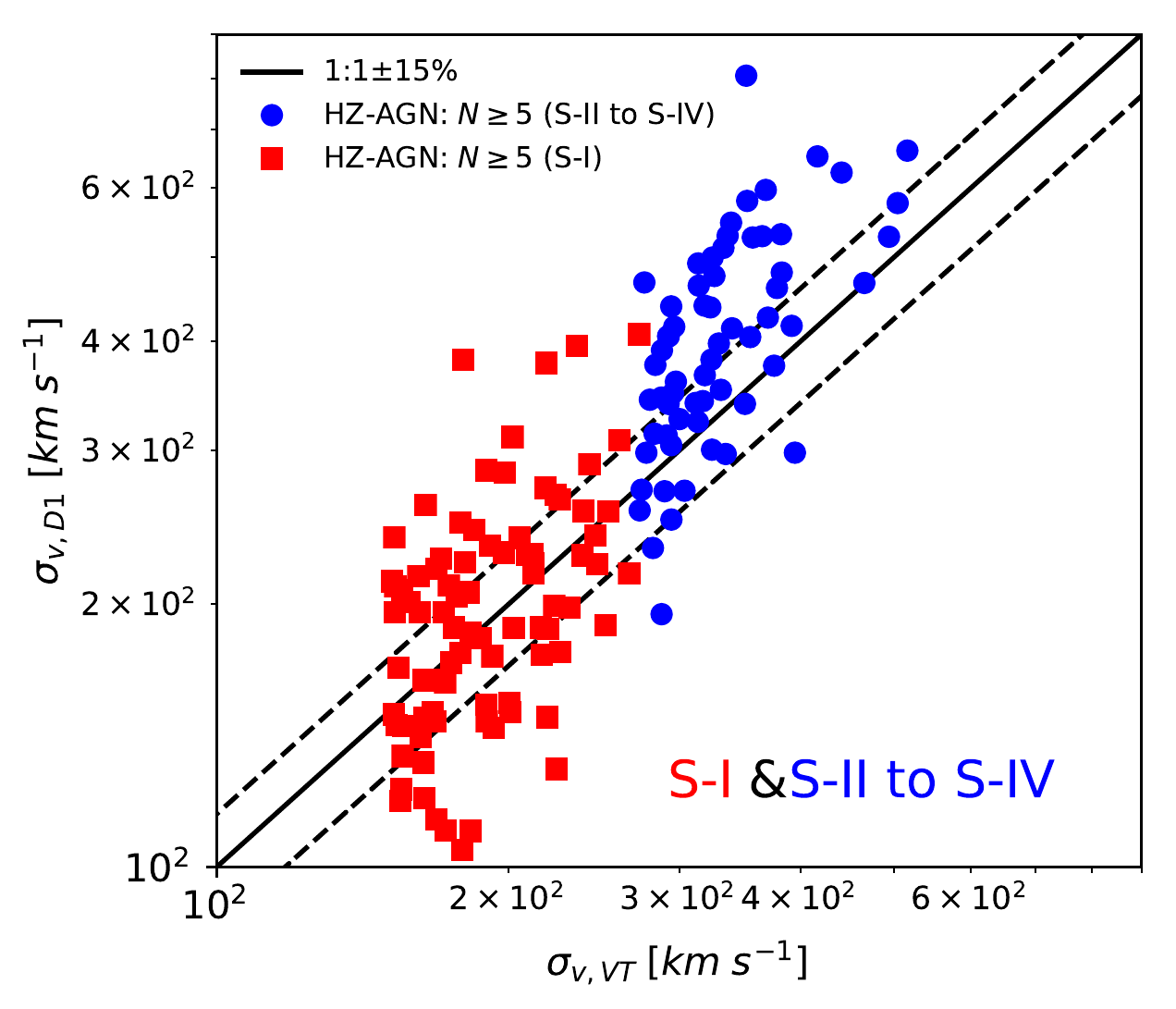}
    \caption{Observed velocity dispersion ($\sigma_{v, \mathrm{obs}}$) of groups determined using their spec-z members within $r_{\mathrm{200c}}$ versus the velocity dispersion predicted by the virial theorem (Eq~\ref{Eq.1}, $\sigma_{v, \mathrm{VT}}$), in COSMOS based on $L_x$- halo mass scaling relation in observations  (\textit{upper panel}) and using the virial mass in the {\sc Horizon-AGN}  simulation (\textit{lower panel}). Groups with $N\geq 5$  are plotted with filled blue circles and red squares. Groups with  $N\geq10$ spectroscopic members are shown with open green circles. The solid and dashed black lines show the 1:1 relation and $\pm15\%$ intervals. The filled red squares and blue circles represent groups in S-I and S-II to S-IV respectively.}
    \label{fig:sigma_v_x}
\end{figure}

\section{Results} \label{Sec:results}
\subsection{Comparison of the observed velocity dispersion to the prediction from virial theorem} \label{SubSec:vel_disp}

Fig.~\ref{fig:sigma_v_x} compares the velocity dispersion of groups inferred from the virial theorem to the observed velocity dispersion measured from spectroscopy in COSMOS (\textit{upper panel}) and in the {\sc Horizon-AGN} simulation (\textit{lower panel}). Red data points correspond to groups within S-I and  blue points to groups in S-II, S-III, and S-IV. 

Although an overall correlation is recovered, both in COSMOS and in {\sc Horizon-AGN}, the observed $\sigma_{v, \mathrm{obs}}$ is found to scatter significantly at a given $\sigma_{v, \mathrm{VT}}$, especially for low mass groups (S-I). This scatter might be driven either by complex substructures within the groups (or more generally by the anisotropy of galaxy spatial distribution) or by infalling galaxies in non-virialised orbits. At higher masses, the measured line-of-sight velocity also deviates from the 1-to-1 relation in both COSMOS and the {\sc Horizon-AGN} simulation.

 \begin{figure*}
  \centering
 \includegraphics[width=0.45\textwidth]{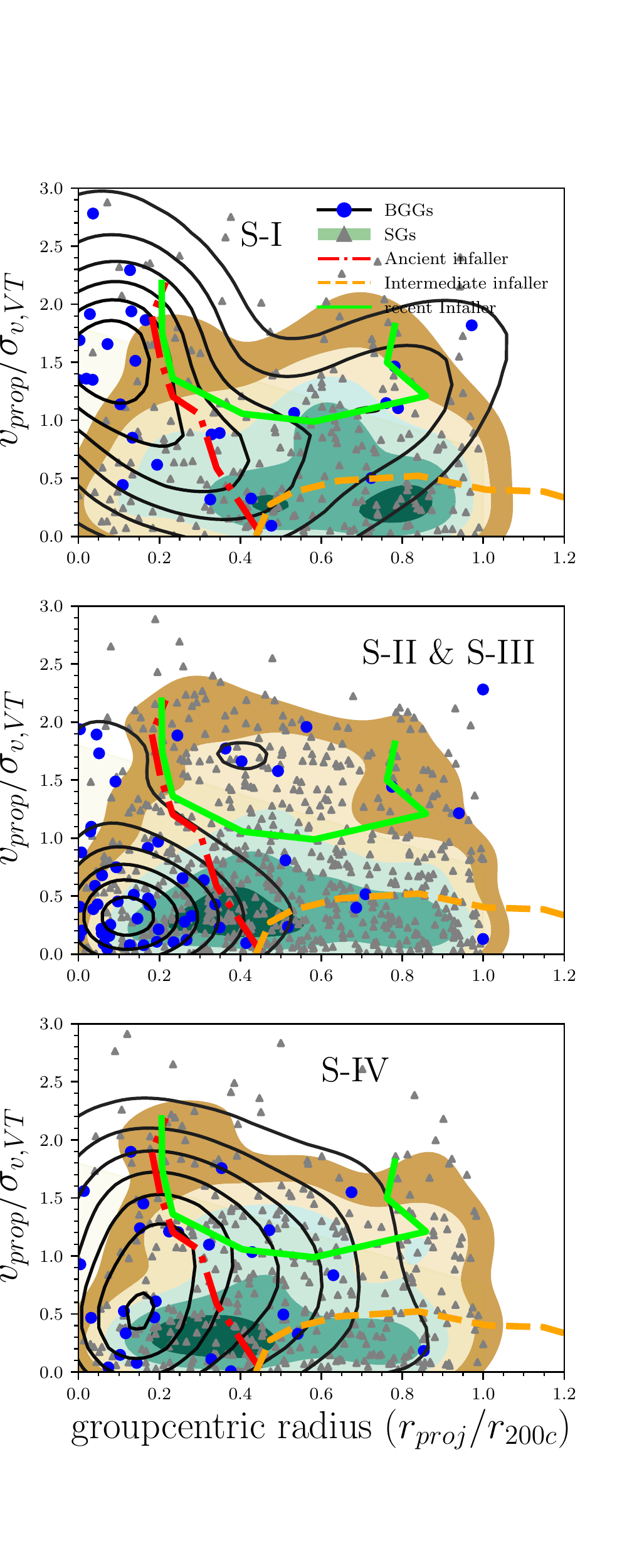}
 \includegraphics[width=0.45\textwidth]{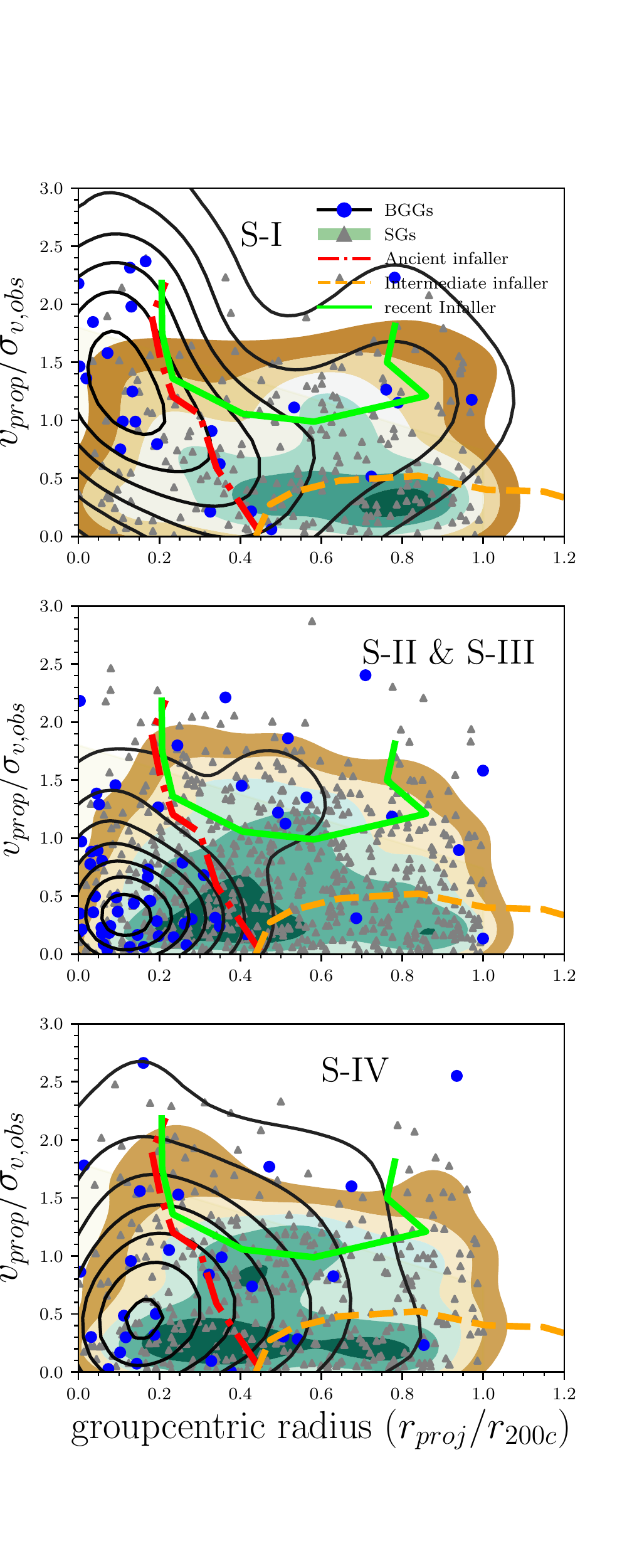}

 \caption{Phase-space diagrams showing the relative line-of-sight velocity of group galaxies as a function of distance from the group X-ray centre for S-I, S-II $\&$ S-III, and S-IV. The orbital velocities of galaxies are normalised either to the velocity dispersion derived using the virial theorem (\textit{left panels}) or to the observed velocity dispersion (\textit{right panels}). The blue and grey data points show the BGGs and all satellite galaxies (SGs) respectively. Solid black lines and shaded area represent the density contours of respectively the BGG and SGs distributions in this plane, estimated using the KDE method. The dash-dotted red, dashed orange, and solid lime isocontours are taken from Fig. 8 in \cite{rhee2017phase} and represent the regions where galaxies in these areas are `ancient infallers', `intermediate infallers', and `recent infallers' with probabilities of 40\%, 25\%, and 40\%. The majority of the BGGs are found in the ancient infallers area. It should be noted however that we probe a lower mass range than described in \cite{rhee2017phase}.} 
  \label{fig:psdiagram}
 \end{figure*}

\subsection{The phase-space diagram of group galaxies} \label{SubSec:phase-space-diagram}

In order to better understand the assembly history of groups and the reason for the scatter observed in Fig.~\ref{fig:sigma_v_x}, we construct phase-space diagrams using the line-of-sight velocity of the member galaxies and the groupcentric radius. The phase-space diagram is used as an indicator of the accretion history of cluster and group member galaxies: galaxies which were recently accreted onto a cluster/group tend to have high relative velocities and large groupcentric radius offsets from the bottom of the potential well (as estimated from the centre of clusters and groups). 

Fig.~\ref{fig:psdiagram} presents the location of group member galaxies in the phase-space diagram for the S-I (\textit{upper panels}), the combined  S-II and  S-III (\textit{middle panels}), and the S-IV sub-samples (\textit{lower panels}).  The proper velocity is normalised either to  $\sigma_{v, \mathrm{VT}}$ (\textit{left panels}) or to  $\sigma_{v, \mathrm{obs}}$ (\textit{right panels}). The BGGs  and satellite galaxies (hereafter, SGs) are shown with filled blue circles and filled grey triangles respectively. 
Solid black lines and shaded areas represent the density contours of respectively the BGG and SGs distributions in this plane, estimated using the Kernel Density Estimation (KDE) method. 
\begin{figure*}
    \centering
    \includegraphics[width=0.495 \textwidth]{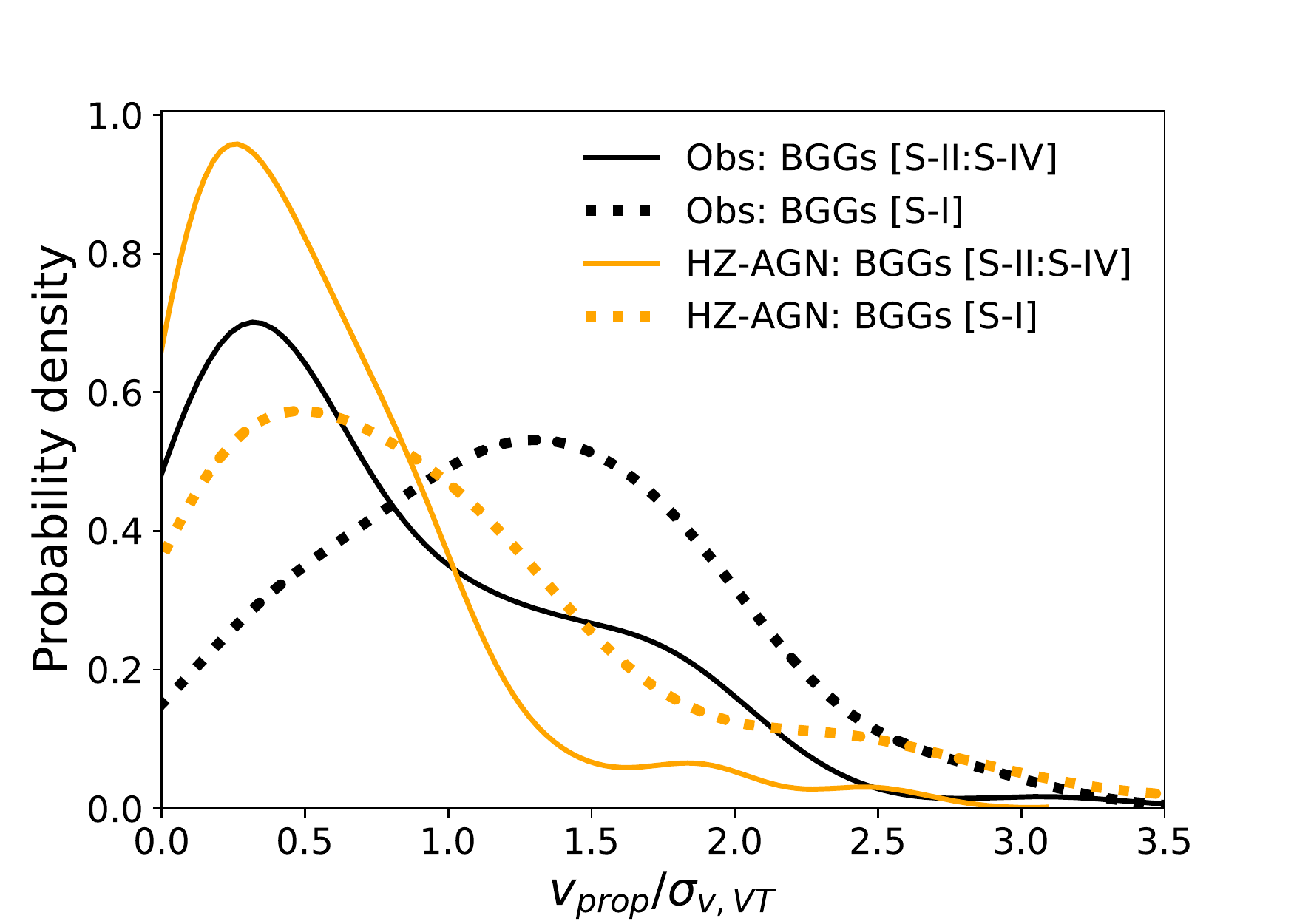}
    \includegraphics[width=0.495 \textwidth]{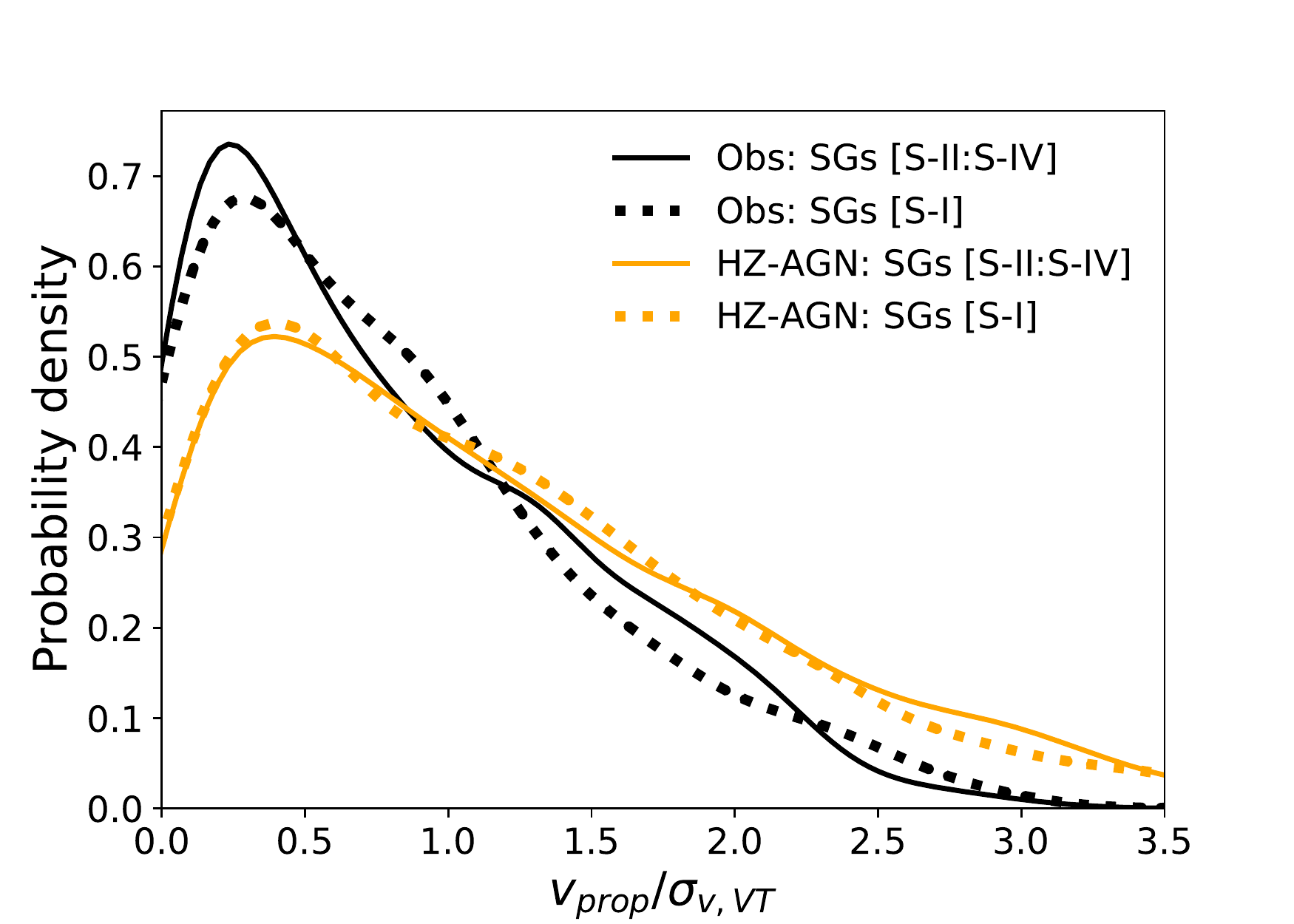}
    \caption{Distributions of the ratio of the line-of-sight velocity to the group velocity dispersion for BGG (\textit{left}) and satellite (\textit{right})  ($v_{\mathrm{prop}}/\sigma_{v, \mathrm{VT}}$), for the S-I (dotted lines) and combined  S-II to S-IV samples (solid lines) in COSMOS (black) and  {\sc Horizon-AGN} (orange). We find no significant redshift evolution of the distribution in both observations and simulation, thus the sub-samples S-II to S-IV are combined here. The distribution for BGGs in low mass groups at $z<0.4$ suggest they are in relatively dynamically unrelaxed systems.}   \label{fig:summ_vlos}
\end{figure*}

Using simulations, \cite{rhee2017phase}  measured for galaxies in this plane the time spent since they crossed the virial radius of the cluster for the first time ($t_{\mathrm{inf}}$) and showed that different locations of galaxies in phase-space correlate with different times since infall ($t_{\mathrm{inf}}$). They subsequently define four different regions in phase-space allowing to classify galaxies as follows: (i)`the first infallers' which have not yet definitively fallen into clusters; (ii)`Recent infallers' whose $t_{\mathrm{inf}}$ ranges as  $0<t_{\mathrm{inf}}/{\rm Gyr}<3.63$; (iii) `intermediate infallers' with  $3.63<t_{\mathrm{inf}}/{\rm Gyr}<6.45$; and  (iv)`ancient infallers', those galaxies having $6.45<t_{\mathrm{inf}}/{\rm Gyr}<13.7$. 
 The highlighted lime, orange, and red isocontours in Fig. \ref{fig:psdiagram} delimit the recent infallers (with probability of 40\%), intermediate infallers (probability of 25\%), and ancient infallers (probability of 40\%) area \citep[see Figure~8 in][]{rhee2017phase}. We note that these regions have been identified for $z=0$ clusters with halo mass of $\sim10^{14}\; M_\odot$. We point out that \cite{rhee2017phase} normalised the projected distance of galaxies from the cluster centre to $R_{\mathrm{vir}}$. We convert this radius to $r_{\mathrm{200c}}$  \citep[see][]{White2001}. 
  \begin{table*}[]
    \caption{The dispersion of the ratio of  the BGGs and SGs line-of-sight velocity  to the group velocity dispersion of in both observations (COSMOS) and {\sc Horizon-AGN}(HZ-AGN). Column~1 presents the sub-sample IDs. Column~2 reports the dispersion of the observed $v_{\mathrm{prop}}/\sigma_{\mathrm{v, obs}}$ of BGGs referred as $\sigma_{\mathrm{BGG,obs}}$. Column~3 presents the dispersion of the $v_{\mathrm{prop}}/\sigma_{\mathrm{v, VT}}$ of BGGs referred as $\sigma_{\mathrm{BGG, VT}}$. As in columns~2 and~3, columns~4 and~5 present the results for SGs. This estimation is performed using the gapper estimator. The Jackknife technique is used to estimate the error. }
    \centering
    \begin{tabular}{ l l l l l }
    \hline\hline\\
sub-sample ID & $\sigma_{\mathrm{BGG, obs}}$&$\sigma_{\mathrm{BGG, VT}}$& $\sigma_{\mathrm{SGs, obs}}$&$\sigma_{\mathrm{SGs, VT}}$\\  \\
       \hline \hline\\
COSMOS (S-I) &1.729 $\pm$ 0.214  &$1.476 \pm 0.128$ &$1.016 \pm 0.051$ &$0.919 \pm 0.040$\\
 COSMOS (S-II) &$0.998 \pm 0.137 $&$1.051 \pm 0.147$&$1.066\pm 0.035$&$0.947 \pm 0.032$\\
 COSMOS (S-III) &$0.893\pm 0.151$ &$0.802\pm 0.123$&$0.966\pm 0.047$ &$0.889 \pm 0.039$\\
 COSMOS (S-IV) &$1.066 \pm 0.188$ &$1.071 \pm0.219$&$1.047 \pm 0.039 $&$0.922 \pm 0.032$\\
 \\
     \hline\\
HZ-AGN (S-I) &$1.350\pm 0.289$& $1.203 \pm 0.167$ &$1.421\pm 0.052$&$1.488\pm 0.044$ \\
HZ-AGN (S-II) &$0.867\pm 0.237$& $0.819\pm 0.270$ &$1.181 \pm 0.048$& $1.214\pm 0.050$ \\
HZ-AGN (S-III)& $0.476\pm 0.106$& $0.589\pm 0.175$& $1.075\pm 0.028$&$ 1.402 \pm 0.036$\\
HZ-AGN (S-IV)& $0.851 \pm 0.155$&$1.116\pm 0.193$& $1.123 \pm 0.036$&$ 1.462\pm 0.048$\\
\\
\hline
\end{tabular}

    \label{tab1:mean_vlos}
\end{table*}

Fig.~\ref{fig:psdiagram} is built of groups with more than 5 spectroscopically identified members, but the results do not change significantly if increasing the minimum number of group members (e.g. $N\geq10$). In addition, no significant difference is found depending on which of the velocity dispersion values (observed or theoretically predicted) are chosen to normalise the  line-of-sight velocity (compare \textit{left} and \textit{right} panels).
 
For low mass groups, the distribution of BGGs in  phase-space is quite extended, with a wide range of both  orbital velocities and projected distances to the X-ray centre of halos. The density map peaks at a relatively high velocity ($v_{\mathrm{prop}}\sim1.5\; \sigma_{v, \mathrm{obs}}$) and is slightly offset from the groups centres ($r_{\mathrm{proj}}/r_{\mathrm{200c}}\sim0.1$). BGGs are mostly ancient infallers according to the classification of  \cite{rhee2017phase}\footnote{It should be noted that \cite{rhee2017phase} have determined the recent, intermediate, and ancient infallers' regions in the phase-space diagram for all groups members without making distinction between satellites and  BCGs.}.

In contrast to the BGGs, the SG distribution of S-I groups follows a different trend. SGs density map peaks at  $0.4<r_{\mathrm{proj}}/r_{\mathrm{200c}}<-0.8$ and below $v_{\mathrm{prop}}<1.0\times \sigma_{v, \mathrm{VT}}$ in the intermediate infallers regions according to the classification from \cite{rhee2017phase}. However, the whole population consists with various types of recent, intermediate and ancient infallers. This distribution illustrates that SGs within low mass groups tend not to have any preferred velocity direction.
In the case of higher mass halos, the peak of the BGG distribution lies below $v_{\mathrm{prop}}\sim1.0 \times \sigma_{v, \mathrm{VT}}$ and at $r_{\mathrm{proj}}/r_{\mathrm{200c}}\sim0.5$. Here again (and as expected), the BGGs mostly  occupy the ancient infaller region.
In summary, we find that the BGGs within low mass groups at $0.07\leq z <0.4$ are kinematically distinct from the bulk of the population of other member galaxies (either satellite galaxies or BGG in higher mass groups).

\subsection{Distribution of the BGG line-of-sight velocities}
The \textit{left panel} of Fig~\ref{fig:summ_vlos} shows the KDE distribution of the ratio of the BGG line-of-sight velocity to $\sigma_{v, \mathrm{VT}}$ for S-II to S-IV (solid black line) and  S-I (dotted black line) and compares them to predictions from {\sc Horizon-AGN}. The \textit{right panel} similarly illustrates the  $v_{\mathrm{prop}}/\sigma_{\mathrm{v, VT}}$ distributions for the SGs in observations and the predicted distributions from  {\sc Horizon-AGN}.

In agreement with predictions from {\sc Horizon-AGN} simulation, the distribution of $v_{\mathrm{prop}}/\sigma_{\mathrm{v, VT}}$ for BGGs within massive groups (S-II to S-IV) peaks at $\sim0.5$, indicating that BGGs in massive halos are well settled at the bottom of the potential well, which is an indication for the group to be relaxed.
BGGs peculiar velocities in low mass groups (S-I) are distributed over a larger dynamical range than in high mass groups. A similar trend is seen in the {\sc Horizon-AGN} simulation, although the observed and predicted distributions have different amplitudes, modes and shapes.

For SGs, the distributions of  $v_{\mathrm{prop}}/\sigma_{\mathrm{v, VT}}$  in the S-I and S-II to S-IV sub-samples, in both observations and the simulation, peak below $v_{\mathrm{prop}}/\sigma_{\mathrm{v, VT}}\sim0.5$. This indicates that there is no preferred direction for the velocity of satellite galaxies, so it is likely that satellite galaxies are isotropically distributed.

Using the gapper estimator \citep{beers90}, we measured the velocity dispersion of the BGG and SG populations associated with each sub-sample (S-I to S-II). Since all groups within each sub-sample do not have the same velocity dispersion, we measured the dispersion of the  normalised and dimensionless line-of-sight velocities expressed as $v_{\mathrm{prop}}/\sigma_{\mathrm{v, obs}}$ and  $v_{\mathrm{prop}}/\sigma_{\mathrm{v, VT}}$ for BGGs within each sub-sample. In Table~\ref{tab1:mean_vlos} we refer to these unitless dispersions as $\sigma_{\mathrm{BGG, obs}}$ and $\sigma_{\mathrm{BGG, VT}}$, and similarly as $\sigma_{\mathrm{SGs, obs}}$ and $\sigma_{\mathrm{SGs, VT}}$ for SGs and we report these values for both observations and the {\sc Horizon-AGN} simulation.

For BGGs in low mass groups (S-I), we measure $\sigma_{\mathrm{BGG, obs}}=1.729 \pm 0.214$ and $\sigma_{\mathrm{BGG, VT}}=1.476 \pm 0.128$) in COSMOS, while we have $\sigma_{\mathrm{BGG, D1}}=1.350 \pm 0.289$  and $\sigma_{\mathrm{BGG, VT}}=1.289 \pm 0.167$ in {\sc Horizon-AGN}. 

For the S-II to S-IV combined samples, we measure a $\sigma_{\mathrm{BGG, obs}}=1.010 \pm 0.087$ and $\sigma_{\mathrm{BGG, VT}}=1.009 \pm 0.089$ in observations and $\sigma_{\mathrm{BGG, D1}}=0.821 \pm 0.120$ and $\sigma_{\mathrm{BGG, VT}}=0.931 \pm 0.114$ in {\sc Horizon-AGN}.

\begin{figure*}
    \centering
    \includegraphics[width=0.7\textwidth]{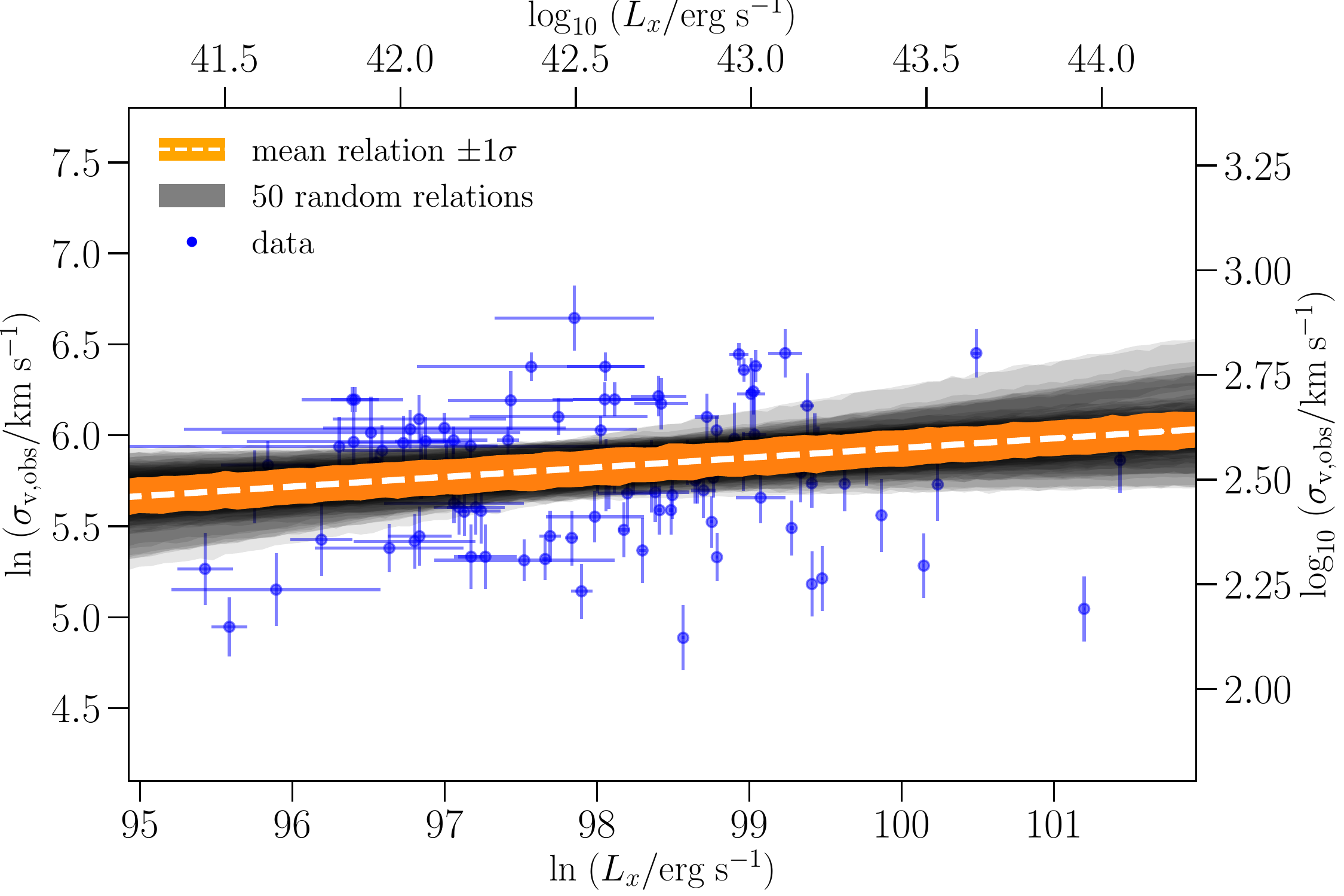}
            \includegraphics[width=0.7\textwidth]{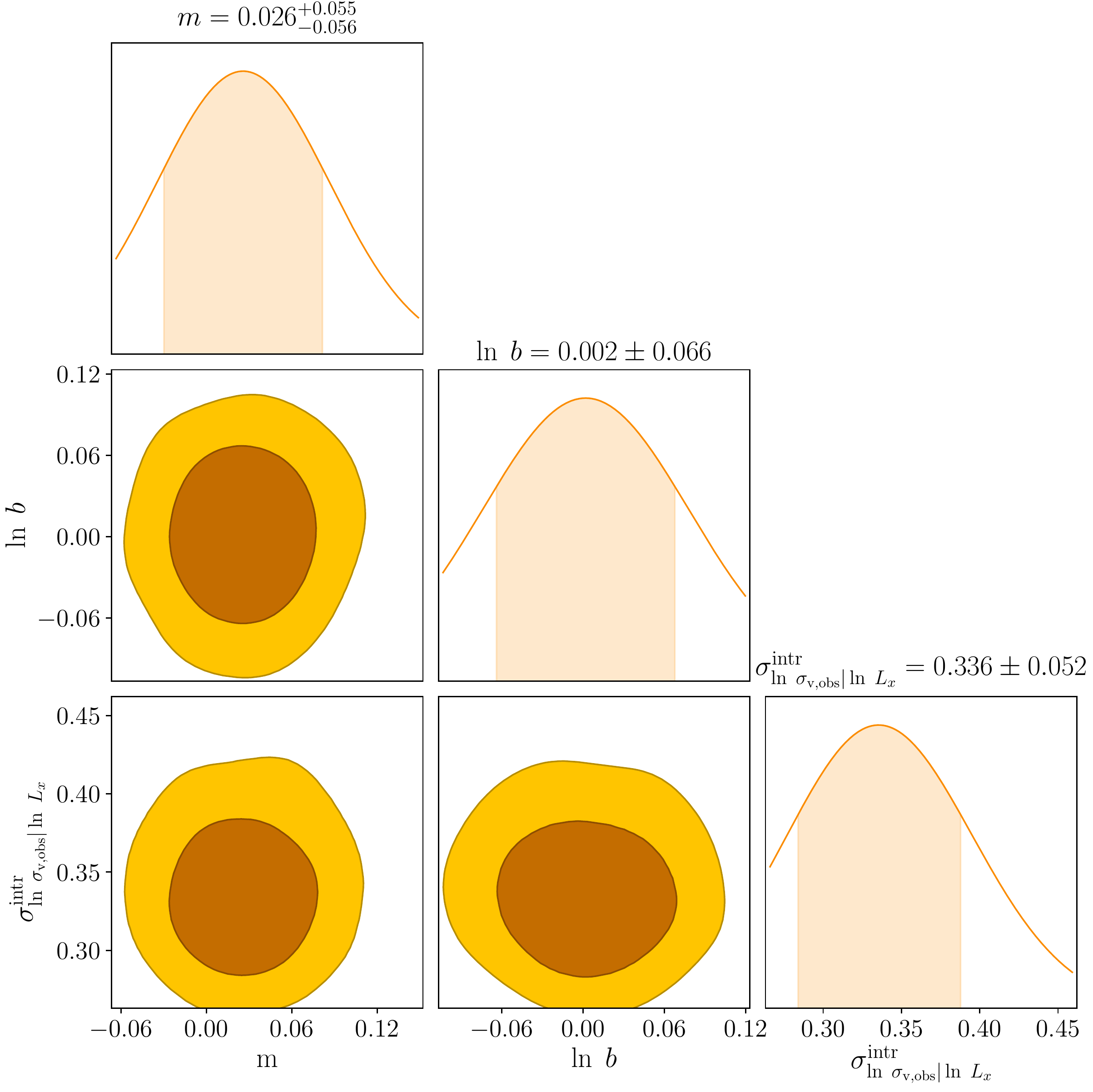}

    \caption{(\textit{Upper panel:}) The scaling relation between the natural logarithm of the observed velocity dispersion of groups ($\ln\; (\sigma_{v, \mathrm{obs}}/km\;s^{-1})$) determined using their spec-z members within $r_{\mathrm{200c}}$ and the natural logarithm of the X-ray luminosity of groups ($\ln\;(L_x/erg\;s^{-1})$) at $z<1.0$ in COSMOS (blue points with associated errors). We also added axes showing the 10-base logarithmic scale. The highlighted orange area shows $\pm1\sigma$ uncertainties around the mean scaling relation (white dashed line). The black lines are a set of 50  different realisations, drawn from the multivariate Gaussian distribution of the parameters ($\left<m\right>=0.026, \left<\ln b\right>=0.002, \left<\sigma^{\mathrm{intr}}_{\ln \sigma_{v, \mathrm{obs}}|\ln L_x}\right>=0.336$) and the scatter covariance matrix is estimated from the MCMC chain.  (\textit{Lower panel:})  The one- and two dimensional marginalised posterior distributions of parameters of the scaling relation (Eq. \ref{Eq.2}), shown as 68\% and 95\% credible regions.}
    \label{fig:sigma_lx_post}
\end{figure*}

\subsection{The X-ray luminosity and velocity dispersion scaling relation} \label{SubSec:Lx-sigma}
The X-ray luminosity-velocity dispersion ($L_X-\sigma_{v,obs}$) relation of galaxy clusters and groups is critical to understanding the dynamical states of clusters and groups and their impact on the scaling relations, as well as the X-ray selection \citep[e.g.][]{Wu1999,Plionis2004,Zhang2011}. Within the context of this work, we aim to quantify how much the offset from the $L_X-\sigma_{v,obs}$ relation is a consequence of kinematic unrest of the group, as quantified from the proper velocity of the BGG. We start by determining the correlation between the X-ray luminosity, $\mathrm{\log}(L_x)$, and velocity dispersion,  $\mathrm\;{\log}\;(\sigma_{\mathrm{v,obs}})$. Using the data shown in Fig.~\ref{fig:sigma_lx_post}, 
we find a positive Pearson correlation coefficient of $r=0.51$.  
In fitting the scaling relation, we normalise the velocity dispersion and  X-ray luminosity of groups and convert them to dimensionless parameters using their median values: $\sigma_{v, \mathrm{pivot}}=315$ km s$^{-1}$ and $L_{x, \mathrm{pivot}}=3.93\times 10^{42}$ erg s$^{-1}$. The relation between $\sigma_{v, \mathrm{obs}}$ and $L_x$ in the natural logarithmic scale  can be approximated  by a power-law as follows: 

\begin{equation}
    \ln\left(\frac{\sigma_{v, \mathrm{obs}}}{\sigma_{v, \mathrm{pivot}}}\right)=  \ln\left[b\times \left(\frac{L_x}{L_{x, \mathrm{pivot}}}\right)^{m}\right],
    \label{Eq.2}
\end{equation}

where m is the slope of the relation and b is the intercept of the relation.  We fitted Eq. \ref{Eq.2} to the data, taking into account the observed uncertainties and an expected intrinsic scatter ($\sigma^{\mathrm{intr}}_{\ln \sigma_{v, \mathrm{obs}}|\ln L_x}$) in the likelihood function being fit. The maximum of likelihood function and the errors on parameters are determined following the procedure presented by \cite{hogg2010}.

The \textit{upper panel} of Fig.~\ref{fig:sigma_lx_post} presents the scaling relation between the observed  $\ln\; (\sigma_{v, \mathrm{obs}}/\mathrm{km\;s^{-1}})$, and $\ln\;(L_x/\mathrm{erg\; s^{-1}})$  for the sample of X-ray groups galaxies in COSMOS (blue circles). We added axes showing the 10-base logarithmic scale too. The dashed white line illustrates the best-fit scaling relation. The orange area represents $\pm 1 \sigma$ errors estimated using the Markov Chain Monte Carlo (MCMC) method \citep[see][]{hogg2010}. The black lines correspond to 50  different realisations of the scaling relation, drawn from the multivariate Gaussian distribution of three parameters (slope, intercept and intrinsic scatter). The \textit{lower panel} represents the one- and two-dimensional projections of the marginalised posterior probability distributions of parameters in
our model (Eq. \ref{Eq.2}) from our re-sampling. The slope and intercept of the mean scaling relation (Eq. \ref{Eq.2}) are found as: $m=0.026^{+0.055}_{-0.056}$, $\ln b=0.002\pm0.066$, and $\sigma^{\mathrm{intr}}_{\ln \sigma_{v, \mathrm{obs}}|\ln L_x}=0.336\pm0.052$.  We note that the error on the estimated parameters corresponds to $\pm 1\sigma$ uncertainty. 

We then try to understand the scatter in this relation by measuring the correlation between the offset from the fitted $L_x-\sigma_{v, \mathrm{obs}}$ relation (hereafter $\Delta\sigma$) and $v_{\mathrm{prop}}/\sigma_{v, \mathrm{obs}}$, which quantifies to some extent the kinematic unrest of the groups. We use S-I and the combined  S-II to S-IV sub-samples, referring to these as low and high mass groups, respectively. Then the linear correlation between this parameter and the BGG proper velocity is estimated using the Pearson correlation coefficient. In the simulation, $\Delta\sigma$ is given as the difference between the one dimensional velocity dispersion and the velocity dispersion measured from the virial theorem ($\Delta\sigma=\sigma_{\mathrm{v,VT}}-\sigma_{\mathrm{v,1D}}$). We assume $\sigma_{\mathrm{v,VT}}$ as the true velocity dispersion of groups. The correlation coefficient between  $\Delta\sigma$ and $v_{\mathrm{prop}}/\sigma_{v, \mathrm{obs}}$ of BGGs within low mass groups in the observations and the {\sc Horizon-AGN} simulation are found to be equal to $0.63\pm0.12$ and $0.36\pm0.03$  in comparison with those for the BGGs in high mass groups, which are  $0.37\pm0.12$ and $-0.07\pm0.03$, respectively. We used the Jackknife method to estimate the error on the correlation coefficients.

The correlation in simulation is found to be less than that in observation. While the influence of gravitational interaction on the velocity of the member galaxies is a result of n-body interaction, the qualitative effect can be estimated using the tidal approximation of \cite{Spitzer1958}. The extra energy acquired by the perturbed object scales with $M \times <r^2>$ (averaging matter density with $r^2$ weight), so in case the total matter profile of the group is either steeper or flatter than $r^{-2}$ density profile, it acquires differences in velocities, which manifest themselves in the sloshing of the core. The larger velocity difference observed could therefore be interpreted as evidence for stronger departures, for instance a larger concentration and a larger scatter in the total matter profiles, compared to that in a simulation. In addition, we cannot rule out the possibility that X-ray selection leads to preferential selection of more concentrated halos, which enhances the effect.
\section{Summary and conclusions}
We construct a phase-space diagram for X-ray galaxy groups in the COSMOS survey using the determinations of group centres based on Chandra imaging and demonstrate that the brightest group galaxies in low mass halos ($<2\times10^{13}\; M_\odot$) are distributed over a wide range of orbital velocities, in contrast to BGGs in more massive groups at similar ($z<0.4$) or higher redshifts. The BGGs in massive groups are more likely to be located at the bottom of the potential well.

We determine the correlation coefficient between the relative proper velocity of the BGGs ($v_{\mathrm{prop}}/\sigma_{v, \mathrm{obs}}$) and the offset from the fitted $L_x-\sigma_{v, \mathrm{obs}}$ scaling relation. We find a positive correlation coefficient of 0.63 for the low mass sub-sample of groups (S-I) in comparison with the lower correlation coefficient of 0.37 for the high mass groups (S-II to S-IV), while the Horizon-AGN simulations are characterised by the values of 0.4 and 0.2, correspondingly.
The uncertainty of the correlation coefficients measurements is found to be 0.12, using a Jackknife resampling method. 
These results argue in favour of the dominant role of group dynamics in the scatter in the $L_x-\sigma_{v, \mathrm{obs}}$ scaling relation. Compared to expectations from numerical simulations based on deviations from the halo mass, a stronger correlation between the relative line-of-sight velocity of BGG and the offset from the $L_x-\sigma_{v, \mathrm{obs}}$  in the observations might be due to an additional (negative) covariant scatter in $L_x$ with the merger state of clusters and groups, as has been studied in detail in \cite{Mulroy2017,Mulroy2019}.

\section{Acknowledgements}
 We acknowledge M. Salvato and G. Hasinger and entire zCOSMOS, DEIMOS, VUDS, MOSFIRE, FORS2, FMOS, MOIRCS, LEGA-C, and the COSMOS teams for their effort in the COSMOS spectroscopic observations and their willingness to share the spectroscopic redshift catalogues.
 
 We wish to thank Kenneth P. K. Quek for his useful comments. 
The author acknowledges the usage of the following python packages, in alphabetical order: \texttt{astropy} \citep{astro2013,astro2018}, \texttt{chainConsumer} \citep{ChainConsumer2019}, 
\texttt{emcee} \citep{emcee3}, \texttt{matplotlib} \citep{mathlib}, \texttt{numpy} \citep{numpy1,numpy2}, and \texttt{scipy} \citep{scipy}.

\end{document}